\documentclass[fleqn,twoside]{article}
\usepackage{espcrc2}

\usepackage{graphicx}
\usepackage{epsfig}
\newcommand{\bel}[1]{\begin{equation}\label{#1}}
\newcommand{\bal}[1]{\begin{eqnarray}\label{#1}}
\newcommand{\ee}{\end{equation}}
\newcommand{\ea}{\end{eqnarray}}
\newcommand{\equ}[1]{~Eq.(\ref{#1})}

\newcommand{\zb}{\bar{z}}

\newcommand{\bmath}{\begin{displaymath}}
\newcommand{\emath}{\end{displaymath}}
\newcommand{\bite}{\begin{itemize}}
\newcommand{\eite}{\end{itemize}}

\newcommand{\eps}{\varepsilon}
\newcommand{\drop}[1]{}

\newcommand{\AmS}{{\protect\the\textfont2
 A\kern-.1667em\lower.5ex\hbox{M}\kern-.125emS}}

\title{Semiclassical Electromagnetic Casimir Self-Energies}

\author{ Martin Schaden\thanks{email: mschaden@andromeda.rutgers.edu}\\
Rutgers University, 211 Smith Hall, 101 Warren St., Newark, NJ 07102, U.S.A.}

\begin{document}

\begin{abstract}
\noindent The electromagnetic Casimir energies of a spherical and a cylindrical cavity are analyzed
semiclassically. The field theoretical self-stress of a spherical cavity with ideal metallic boundary conditions
is reproduced to better than 1\%. The subtractions in this case are unambiguous and the good agreement is shows
that finite contributions from the exterior of the cavity are small. The semiclassical electromagnetic Casimir
energy of a cylindrical cavity on the other hand vanishes to any order in the \emph{real} reflection
coefficients, whereas the field theoretic Casimir energy of a cylindrical boundary that is perfectly metallic
and infinitesimally thin is finite and negative\cite{DRM81}. Contrary to the spherical case and in agreement
with Barton's perturbative analysis\cite{Barton01}, the subtractions in the spectral density for the cylinder
are not universal when only the interior modes of are taken into account\cite{Saharian00}. The Casimir energy of
a cylindrical cavity therefore depends sensitively on the physical nature of the boundary in the ultraviolet
whereas the Casimir energy of a spherical one does not.  The extension of the semiclassical approach to more
realistic systems is sketched.  \vspace{1pc}
\end{abstract}


\maketitle

\section{INTRODUCTION}
\label{Intro} Demonstrating that the collective interaction of atomic systems may have macroscopic consequences,
Casimir obtained the now famous attractive force between two neutral metallic plates\cite{Casimir48} from the
boundary conditions they impose on the electromagnetic field. Half a century later, his prediction has been
verified experimentally\cite{Experiments99} to better than 1\%.

Twenty years after Casimir's result for two parallel plates, Boyer calculated the zero-point energy of an ideal
conducting spherical shell\cite{Boyer68}. Contrary to intuition derived from the attraction between two parallel
plates, the spherical shell tends to expand. Boyer's result has since been improved in accuracy and verified by
a number of field theoretic methods\cite{Davies72,BD78,Milton78,LNB99,EKK00} -- even though there may be little
hope of observing this effect experimentally in the near future\cite{Barton04}.

Since field theoretic methods require explicit or implicit knowledge of cavity \emph{frequencies}, they have
predominantly  been successfully employed to obtain the Casimir energies of classically \emph{integrable}
systems. Thus, in addition to a spherical cavity, the electromagnetic Casimir energies of dielectric
slabs\cite{Lifshitz56,Milloni93,Milton04}, metallic parallelepipeds\cite{BB70,Lukosz71,DC76,AW83,Edery06} and
long cylinders\cite{LNB99,DRM81,GR98,BN94,MNN99,KR00,NP00} have been computed in this manner.

However, most systems are not integrable and often cannot even be approximated by such systems. It thus is
desirable to develop reliable methods for estimating the Casimir energies of classically non-integrable and even
chaotic systems. Balian, Bloch and Duplantier calculate Casimir energies based on a multiple scattering
expansion to the Green's function\cite{BD78,BB70,BD04}. This approach does not require knowledge of the quantum
mechanical spectrum and the geometric expansion is exact for sufficiently smooth and ideally metallic cavities.
Ultra-violet divergent contributions have to be subtracted at every order and the relative importance of the
finite remainders is hard to assess a priori. It in practice is often difficult to carry the (in principle
exact) expansion beyond the first few terms without semiclassical approximation\footnote{Restricting to just two
reflections, the Casimir energy of an idealized spherical cavity of radius $R$ is of the correct sign and quite
accurately estimated\cite{BD78} as $3\hbar c/(64 R)\sim 0.0469\hbar c/R$ }. In\cite{MS98} a semiclassical method
based on Gutzwiller's trace formula\cite{Gutzwiller90} for the response function was proposed to estimate
(finite) Casimir energies. It is appropriate for Casimir energies of hyperbolic and chaotic
systems\cite{MS98,SS05,BMW06} with isolated classical periodic orbits. The semiclassical approximation to the
oscillating part of the spectral density of integrable systems was derived in\cite{BB70,BT76} and has been used
to estimate the force between concentric metallic cylindrical shells\cite{MSSS03}. The agreement with exact
results is rather impressive even when the radius of the outer cylinder is about four times that of the inner
one\cite{MSSS03}. Although not exact in general, the semiclassical approximation associates the finite (Casimir)
part of the vacuum energy with optical properties of the system. It captures aspects of Casimir energies that
have been puzzling for some time\cite{Schaden06} and appears to be a reasonable approximation in the
experimentally accessible regime where these forces are relatively large\cite{MS98,Schaden06}. Path integral
methods\cite{Schubert02,GLM03,GK05,EHGK01} in principle allow one to obtain Casimir interactions between
disjoint bodies to arbitrary precision. Due to unresolved renormalization issues, these methods have so far not
been used to study the self stress of cavities. The purpose of this article is to revisit and analyze the
Casimir stress of (integrable) spherical and cylindrical cavities semiclassically and to compare it with field
theoretic results.

The simplicity, transparency and somewhat surprising accuracy of the semiclassical approximation is demonstrated
in Boyer's problem\cite{Boyer68,Davies72,BD78,Milton78,LNB99,EKK00}, that is in determining the electromagnetic
Casimir energy of a spherical cavity with an (ideal) metallic boundary. Instead of directly proceeding from the
semiclassical expressions for the oscillating part of the spectral density\cite{BB70,BT76}, our starting point
will be the dual representation of the Casimir energy of integrable systems in terms of periodic
orbits\cite{Gutzwiller90,BT76}. The subtractions in the spectral density that render the Casimir energy finite
then are apparent and we can better analyze the semiclassical approximation. Even so, the semiclassical analysis
of Boyer's problem is an order of magnitude simpler than any given previously. However, since no bounds are
obtained, it at present is not possible to judge the accuracy of the result without comparing to field theoretic
calculations\cite{Milton78}. It will become rather clear though, that the semiclassical analysis is accurate
enough to infer the sign of the Casimir energy of a cavity by geometric arguments if the contribution from
periodic orbits does not vanish.

In ref.\cite{MSSS03} periodic orbits in fact were found not to contribute to the Casimir energy of a long
cylindrical cavity. The optical phases that give a positive semiclassical Casimir energy for the spherical
cavity, lead to a vanishing one in the cylindrical case. The somewhat intriguing result that the Casimir energy
of a cylinder\cite{LNB99,MNN99,BD04,Barton01,CM05,RM05} vanishes to first order in the reflection coefficients
thus apparently is readily explained by geometric optics. However, the semiclassical Casimir energy of a
cylindrical cavity vanishes to all orders in the \emph{real} reflection coefficients and thus also vanishes for
an ideal metallic cavity. The discrepancy to the finite field theoretic Casimir energy of an idealized,
infinitesimally thin cylindrical boundary between non-dispersive media with the same speed of
light\cite{DRM81,GR98,MNN99} will be traced to the presence of a logarithmic divergence observed by
Barton\cite{Barton01} in his perturbative treatment of the non-ideal dilute case. We will argue that the exact
cancellation of this divergence in the field theoretic approach is due to the infinitesimal thickness of the
boundary -- interior and exterior contributions to the Casimir energy in this case depend on only one common
scale -- the radius of the cylinder.

\section{The Dual Picture: Casimir Energies in Terms of Periodic Rays}
Integrable systems may be semiclassically quantized in terms of periodic paths on invariant
tori\cite{Einstein17} -- in much the same manner as Bohr first quantized the hydrogen atom. Although in general
not an exact transformation, classical periodic orbits on the invariant tori are \emph{dual} to the mode
frequencies in the semiclassical sense. Applying Poisson's summation formula, the semiclassical Casimir energy
(SCE) due to a massless scalar may be written in terms of classical periodic
orbits\cite{Gutzwiller90,BT76,Schaden06},
\bal{dualtrafo}
{\cal E}_c &=& {\frac1 2}\sum_{\bf n} \hbar\omega_{\bf n}\ - \ {\rm UV~subtractions} \nonumber\\
&\sim & \ \frac{ 1}{ 2 \hbar^d}{\sum_{\bf m}}^\prime e^{-\frac{i \pi}{2}\beta_{\bf m}} \int_{sp} {\bf
dI}\, H({\bf I})\, e^{2\pi i\, {\bf m}\cdot{\bf I}/\hbar}\ .\nonumber\\
\ea
The components of the $d$-dimensional vector ${\bf I}$ in\equ{dualtrafo} are the actions of a set of properly
normalized action-angle variables that describe the integrable system. The exponent of the integrand
in\equ{dualtrafo} is the classical action (in units of $\hbar$) of a periodic orbit that winds $m_i$ times about
the $i$-th cycle of the invariant torus.  $H({\bf I})$ is the associated classical energy and $\beta_{\bf m}$ is
the Keller-Maslov index\cite{Keller58,Maslov72} of a class of periodic orbits identified by the vector of
integers ${\bf m}$. The latter is a topological quantity that does not depend on the actions ${\bf I}$. To
leading semiclassical order, the (primed) sum extends only over those sectors ${\bf m}$ with classical periodic
paths of finite action (see below). The correspondence in \equ{dualtrafo} can only be argued
semiclassically\cite{Gutzwiller90,BT76} and the integrals on the RHS therefore should be evaluated in stationary
phase approximation ($sp$).

Contributions to the Casimir energy from high frequencies correspond to those from short periodic orbits in this
dual picture. Divergences due to periodic classical paths of vanishing length (and thus vanishing action) on the
RHS of\equ{dualtrafo} are related to ultra-violet divergences of the mode sum on the LHS of \equ{dualtrafo}. If
these divergences can be subtracted unambiguously\cite{Schaden06,Barton01,CD79}, the dependence of the vacuum
energy on macroscopic properties of the system is semiclassically represented by contributions due to classical
periodic orbits of finite action only. The primed sum on the RHS of\equ{dualtrafo} indicates this
restriction\footnote{This is conceptually not so different from considering only the contribution of
topologically non-trivial "instanton" sectors to the vacuum energy of a field theory.}. The (divergent) Weyl
contribution to the vacuum energy from the ${\bf m}=(0,\dots,0)$-sector in particular has to be subtracted.
Together with an evaluation of the integrals in stationary phase, this defines the semiclassical Casimir energy
(SCE). To physically interpret the SCE, one has to consider the implicit subtractions in the spectral
density\cite{BD04,Schaden06,Barton01} in greater detail.

\section{The Spherical Cavity}
The semiclassical spectrum of a massless scalar is exact for a number of manifolds without
boundary\cite{Dowker71} and the definition of the SCE by the RHS of \equ{dualtrafo} coincides with the Casimir
energy of zeta-function regularization in these cases. It also is exact for massless scalar fields satisfying
periodic-, Neumann- or Dirichlet- boundary conditions on parallelepipeds\cite{Lukosz71,AW83,Schaden06} as well
as for some tessellations of spheres\cite{Schaden06,CD93,Dowker05}. In \cite{MS98} the semiclassical
approximation was argued to give the leading asymptotic behavior of the Casimir energy whenever the latter
diverges as the ratio of two relevant lengths vanishes. All these criteria do not apply to the Casimir
self-stress of a spherical cavity first considered by Boyer\cite{Boyer68}. The latter is an integrable system,
but the semiclassical spectrum is only asymptotically correct\cite{BB70}. There furthermore is no ratio of
lengths in which one might hope to obtain an asymptotic expansion. One therefore cannot expect the semiclassical
approximation to be exact in this case. It nevertheless turns out to be quite accurate.  The SCE is obtained by
performing the integrals of\equ{dualtrafo} in stationary phase and has a very transparent interpretation in
terms of periodic orbits \emph{within} the cavity only. The sign of the SCE of a spherical cavity in particular
will be quite trivially established and the good agreement supports the conjecture that the contribution from
exterior modes mainly serves to cancel the ultra-violet divergence of the contribution from interior modes in
the field theoretic approach\cite{MB94,Saharian00}. The observed discrepancy of 1\% compared to the
field-theoretic result probably can be attributed to the error in the semiclassical estimate of low-lying
eigenvalues of the Laplace operator -- which is of similar size. Since boundary conditions for the
electromagnetic field never are ideal, many corrections of similar and even greater magnitude would be required
in any realistic situation.

The electromagnetic Casimir energy of any closed cavity with a smooth and perfectly metallic boundary may be
decomposed into the contribution from two massless scalar fields -- one satisfying Dirichlet's, the other
satisfying Neumann's boundary condition on the surface\cite{BD04}. Because the surface is ideally metallic,
contributions to the spectral density from arbitrarily short closed paths that reflect off (either side of) the
smooth surface cancel each other. Semiclassically there is no (potentially divergent) local contribution to the
Casimir energy from such an idealized surface in the electromagnetic case -- the local surface tension in fact
vanishes for an infinitesimally thin surface\cite{BD04,BGH03}. Note that this cancellation is quite special for
the electromagnetic field and metallic boundaries. It in general does not occur for a massless scalar field
satisfying Neumann or Dirichlet boundary conditions on a spherical surface of arbitrary thickness or in even
dimensions\cite{MB94,BKV99,GJKQSW02,Milton02}. However, it was shown in \cite{MB94} that the Casimir energy due
to a scalar is finite for infinitesimally thin spherical cavities in odd dimensional spaces. The following
argument illustrates the absence of ultraviolet divergent contributions proportional to the "area" of any
\emph{infinitesimally thin} even-dimensional surface if the speed of light in the surrounding medium is
constant. Barring other scales and with a dimensionless cutoff\footnote{In dimensional regularization this
cutoff is just the (small) deviation from integer $d$ and in the semiclassical context it is the ratio of the
cutoff in the length of the shortest optical paths to some characteristic (fixed) length in the problem.}, the
local contribution to the surface divergence from a small ($(d-1)$-dimensional) surface element $dA$ is
proportional to $\hbar c \,dA \sum_i f_i(\{R_j/R_k\})\,/R_i^d $, where $R_i$ are (local) radii of curvature and
the $f_i$ are dimensionless functions of their ratios only.  The surface divergence semiclassically arises due
to arbitrary short closed optical paths on either side of the infinitesimally thin boundary that touch the
surface at a single point only. But the radii of curvature on either side of a point on the infinitesimally thin
surface are of equal magnitude and of opposite sign. This implies that divergent contributions from the two
arbitrarily short closed optical paths with a common point on the surface that lie on either of its sides cancel
precisely for odd $d$, whereas they are equal in even $d$ -- provided the boundary conditions and speed of light
are the same for {\it both } sides of the surface. One therefore expects the scalar Casimir energy of such
cavities to be analytic in the vicinity of odd $d$, but to in general diverge for even dimensional spaces. The
finite Casimir energy of thin spherical cavities in $d$ dimensions\cite{MB94} as well as the perturbative
analysis of such surface divergences\cite{Milton02} agree with this semiclassical analysis. In three dimensions,
the Casimir energy due to a scalar field was found to be finite\cite{GR98,Saharian00,ST06} also for an
infinitesimally thin cylindrical cavity. The above argument is quite general and indicates that surface
divergences in will cancel locally only for \emph{infinitesimally thin} (and sufficiently smooth)
even-dimensional boundaries when the speed of light is the same on either side. This cancellation of surface
divergences is due to the \emph{infinitesimal thinness} of the boundary and does not depend on the precise
nature of the (frequency-independent) boundary condition. [It is quite different from the semiclassical
cancellation of surface divergences of the electromagnetic field on ideal metallic boundaries in three
dimensions. The latter occurs due to the different boundary conditions for the two scalar modes and does not
depend on the thickness of the metallic surface.]

The only subtraction in the spectral density required for a finite Casimir energy in the electromagnetic case
with idealized metallic boundary conditions thus is the Weyl contribution proportional to the volume of the
sphere. The latter corresponds to ignoring the ${\bf m}=(0,0,0)$ contribution to the sum in\equ{dualtrafo}.  The
remaining difficulty in calculating the SCE of an integrable system is a convenient choice of action-angle
variables. For a massless scalar in three dimensions satisfying boundary conditions with spherical symmetry, an
obvious set of actions is the magnitude of angular momentum, $I_2=L$, one of the components of angular momentum
$I_3=L_z$ and an action $I_1$ associated with the radial degree of freedom.

Since the azimuthal angle of any classical orbit is constant, the energy $E=H(I_1,I_2)$ of a massless particle
in a spherical cavity of radius $R$ does not depend on $I_3=L_z$. In terms of the previous choice of actions,
the classical energy is implicitly given by,
\bel{H}
\pi I_1 + I_2 \arccos\left(\frac{c I_2}{E R}\right)=\frac{E R}{c}
\sqrt{1-\left(\frac{c I_2}{E R}\right)^2}\ .
\ee
The  branches of the square root and inverse cosine in \equ{H} are chosen so that $I_1$ is positive. It is
convenient to introduce dimensionless variables
\bel{var}
\lambda=2 E R/(\hbar c)\ \ {\rm and}\ \ z= c I_2/(E R),
\ee
for the total energy  (in units of $\hbar c/(2R)$) and the angular momentum (in units of $E R/c$) of an orbit.
Note that $z\in [0,1]$ and that the semiclassical regime formally corresponds to $\lambda\gg 1$, i.e. to
wavelengths that are much shorter than the dimensions of the cavity. With the help of\equ{H} and the definitions
of\equ{var}, the semiclassical expression in\equ{dualtrafo} for the Casimir energy of a massless scalar field
satisfying Neumann or Dirichlet boundary conditions on a spherical surface becomes,
\bal{sphere}
{\cal E}&=&\frac{\hbar c}{4\pi R}{\sum_{m,n\geq 0}}^\prime
\Re\left[e^{-i\frac{\pi}{2}\beta(n,m)}\times\right.\nonumber\\
&&\hspace{-5em}\left.\times  \int_0^\infty
\hspace{-1em}d\lambda\lambda^3\hspace{-.5em}\int_0^1
\hspace{-.5em}dz z {\scriptstyle \sqrt{1-z^2}}\; e^{i\lambda[n
(\sqrt{1-z^2}-z\arccos(z))+m\pi z]}\right]\
.\nonumber\\
\ea
The integral over $I_3$ has here been performed in stationary phase approximation. Because the Hamiltonian does
not depend on $I_3$, only periodic orbits with $m_3=0$ contribute. Since $-I_2\leq I_3\leq I_2$, one has that
$\int dI_3= 2I_2=\hbar\lambda z$. The factor $2 I_2$ accounts for the $2(l+1/2)$-degeneracy of states with
angular momentum $L=\hbar(l+1/2)=I_2$]. By taking (4 times) the real part in\equ{sphere} one can restrict the
summations to non-negative integers and choose the positive branch of the square root- and inverse cosine-
functions in the exponent\footnote{The primed sum now implies half the summand if one of the integers vanishes
as well as  the absence of the $m=n=0$ term.}. The Keller-Maslov index $\beta(n,m)$ of a classical sector
depends on whether Neumann or Dirichlet boundary conditions are satisfied on the spherical shell and will
shortly be determined.

For positive integers $m$ and $n$, the phase of the integrand in\equ{sphere} is stationary at
$z=\zb(n,m)\in[0,1]$ where,
\bal{sp}
0&=&-n\arccos(\zb)+m\pi\nonumber\\
&&\hspace{-2em}\Rightarrow \ \zb(n,m)=\cos(m\pi/n),\ n\ge 2 m>1\ .
\ea
Restrictions on the values of $m$ and $n$ arise because $\arccos(\zb)\in[0,\pi/2]$ on the chosen branch. The
phase is stationary at classically allowed points only for sectors with $n\ge 2 m>1$. Semiclassical
contributions to the integrals of other sectors arise due to the endpoints of the $z$-integration at $z=0$ and
$z=1$ only. Such "diffractive" contributions are of sub-leading order in the asymptotic expansion of the
spectral density for large $\lambda$. Note that $m\rightarrow m+n$ amounts to the choice of another branch of
the inverse cosine.

The classical action in sectors with stationary points is,
\bal{Scl}
S_{cl}(n,m)&=&\hbar \lambda n
\sin(m\pi/n)\\
&&\hspace{-4em}=(E/c) 2 n R\sin(m\pi/n)=(E/c)L(n,m)\ ,\nonumber
\ea
where $L(n,m)$ is the total length of the classical orbit. Some of these periodic orbits are shown in Fig.~1.
The integer $m$ in\equ{Scl} gives the number of times an orbit circles the origin. The integer $n>1$ gives the
number of times an orbit touches the spherical surface. As indicated in Fig.~1, the set of classical periodic
orbits in the $(n,m)$-sector form a caustic surface and a double covering is required for a unique phase-space
description\cite{Keller58}. The two sheets are joined at the inner caustic [indicated by a dashed circle in
Fig.~1] and at the outer spherical shell of radius $R$. Every orbit that passes the spherical shell $n$ times
also passes through the caustic $n$ times. The cross-section of a bundle of rays is reduced to a point at the
spherical caustic surface. The caustic thus is of second order and is associated with a phase loss of $\pi$
every time it is crossed. At each specular reflection off the outer shell, Dirichlet boundary conditions require
an additional phase loss of $\pi$  whereas there is no change in phase for Neumann boundary conditions.
Altogether the Keller-Maslov index of sector $(n,m)$ depends on $n$ only and is given by,
\bel{KM}
\beta(n,m)=\left\{\begin{array}{rl} 0, &\mathrm{for~Dirichlet~b.c.}\\
2n, &\mathrm{for~Neumann~b.c.}\end{array}\right. \ .
\ee

Since the electromagnetic Casimir energy of a cavity with a smooth surface on which (ideal) metallic boundary
conditions hold can be viewed as due to two massless scalar fields, one satisfying Dirichlet and the other
Neumann boundary conditions\cite{BD04} on this surface, only sectors $(n,m)$ with \emph{even} $n=2 k\ge 2 m\ge
2$ contribute\cite{BD04} to the SCE in leading order of the asymptotic expansion for large $\lambda$.

\hspace{-2em}\epsfig{file=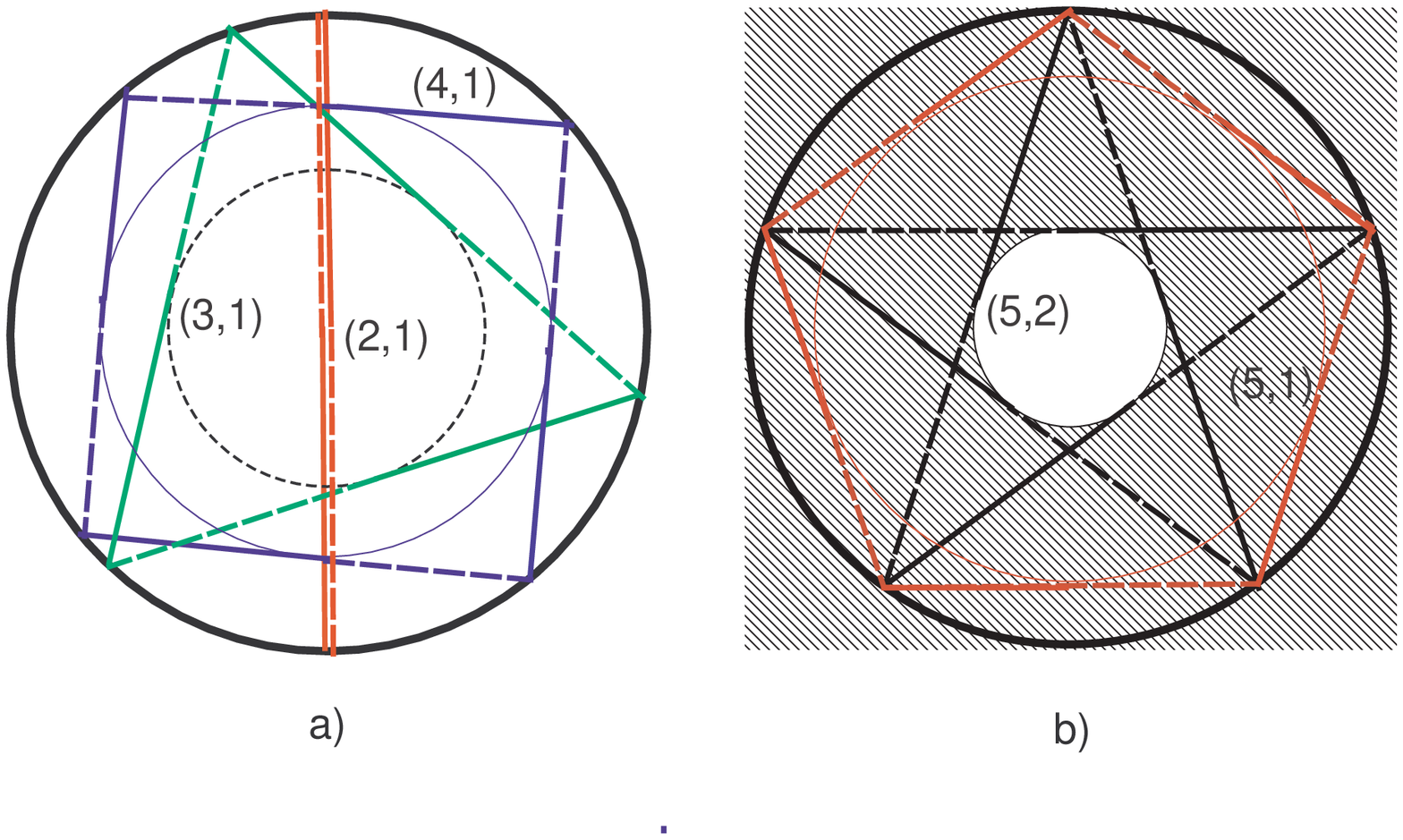,width=3truein}
{\newline\small Fig.1: Classical periodic rays within a ball
or a solid cylinder. a) The shortest primitive rays with winding numbers $(n,m)\in\{(2,1),(3,1),(4,1)\}$. b)
Primitive rays to winding numbers $(n,m)=(5,1)$ and $(5,2)$. Caustic surfaces are shown as thin circles. The
dashed part of any trajectory is on one sheet and its solid part on the other of a two-sheeted covering space.
The "phase space" of the $(5,2)$ sector is indicated by the hatched area. Note that caustics are of $2^{{\rm
nd}}$ order for a spherical cavity but of $1^{{\rm st}}$ order for a cylindrical one. } \vspace{.5cm}

Note that sectors with $m=0$ or $n=0$ have vanishing classical action and do not contribute to the SCE. \equ{sp}
implies that extremal paths in the $(n>0,m=0)$ sectors have maximal angular momentum $1=\zb=l c/(E R)$. These
are great circles that are wholly within the spherical shell in a plane perpendicular to the one under
consideration, i.e. classical orbits with $I_3=L_z=0$. Because the measure of the $z$-integral vanishes at $z=1$
like $\sqrt{1-z}$ these classical paths are extremal but not stationary. This can also be seen by expanding the
exponent in\equ{sphere} about the stationary point $\zb(n,m)$. For $(n>0,m>0)$ the second derivative of the
classical action is finite at $\zb(n,m)$,
\bel{Sdd}
\frac{\partial^2}{\partial\zb^2}[n({\scriptstyle
\sqrt{1-\zb^2}}-\zb\arccos(\zb))+m\pi\zb]=\frac{n}{\sin({\frac{m\pi} n})}\ ,
\ee
whereas it diverges in sectors with $m=0$. The behavior of the exponent for $z\sim 1$ in this case is,
\bel{m0}
{\scriptstyle \sqrt{1-z^2}}-z\arccos(z)={\scriptstyle \frac{2\sqrt{2}}{3}(1-z)^{3/2}}+ {\cal O}({\scriptstyle
(1-z)^{5/2}})\ .
\ee
Quadratic fluctuations about the classical orbit with $m=0$ thus have vanishing width and these sectors do not
contribute in stationary phase approximation. To leading semiclassical accuracy, the Casimir energy of a
spherical cavity with an ideal metallic boundary therefore is,
\bal{sphere1}
{\cal E}^{\rm ball}_{\rm EM}&\sim&\frac{\hbar c}{4\pi R}\,{\rm Re}\sum_{n=1}^\infty
(1^n+(-1)^n)\sum_{m=1}^{n/2} \nonumber\\
&&\hspace{-3.5em}\times  \int_0^\infty \hspace{-1em}d\lambda\lambda^3e^{i n \lambda\sin({\frac{m\pi}
n})}\hspace{-.5em}\int_0^1 \hspace{-.5em}dz
z{\scriptstyle \sqrt{1-z^2}} \; e^{i \frac{n\lambda(z-\zb(n,m))^2}{2\sin(\frac{m\pi}{n})}}\nonumber\\
&&\hspace{-3.5em}\sim\frac{\hbar c}{R}\left[\sum_{k=1}^\infty \frac{1}{16\pi k^4} +\sum_{k=2}^\infty\frac{15
\sqrt{2}}{256 k^4} \sum_{m=1}^{k-1}\frac{\cos(\frac{m\pi}{2k})}{
\sin^2(\frac{m\pi}{2k})}\right]\nonumber\\
&&\hspace{-3.5em}\sim 0.04668...\frac{\hbar c}{R}\ .
\ea

This semiclassical estimate is only about 1\% larger than the best numerical value\cite{Milton78}
$0.04617...\hbar c/R$ for the electromagnetic Casimir energy of a spherical cavity with an infinitesimally thin
metallic surface. The error is of the same order as that of limiting the multiple reflection expansion to just
two reflections, which gives\cite{BD78} ${\cal E}^{\rm ball}_{\rm EM}\sim 0.0469\hbar c/R$.  Note that the
contribution from the $(2k,k)$ sectors had to be considered separately in\equ{sphere1} since the measure $dz z$
vanishes at the stationary point $\zb(2k,k)=\cos(\pi/2)=0$ of the integrand, which is an endpoint of the
integration domain. As can be seen in Fig.~1a), the classical rays of $(2k,k)$-sectors go back and forth between
antipodes of the cavity and pass through its center -- they have angular momentum $L=0=\zb(2 k,k)$.

The shortest primitive orbits give somewhat less than half ($1/(16\pi)\sim 0.02$) of the total SCE of the
spherical cavity -- much less than the 92\% they contribute to the Casimir energy of parallel plates. The main
reason is that contributions only drop off as $1/k^2$ rather than like $1/k^4$ as for parallel plates. The
length of an orbit in the $(4,1)$-sector (the inscribed square in Fig.~1a) furthermore is just a factor of
$\sqrt{2}$ longer than a $(2,1)$-orbit [which in turn is a factor of $1/\sqrt{2}$ shorter than a $(4,2)$-orbit].
To estimate the magnitude of the contribution from any particular sector one has to take the available phase
space as well as the ray's length into account. Thus, although the length of a $(2 k,1)$-orbit tends to $2\pi R$
for $k\rightarrow \infty$, the associated phase-space (essentially given by the volume of the shell between the
boundary of the cavity and the inner caustic) decreases like $1/k^2$.  This accounts for the relatively slow
convergence of the sum in\equ{sphere1}. To achieve an accuracy of $10^{-5}$, the first 50 terms of the sum were
evaluated explicitly  and the remaining contribution was estimated using Richardson's extrapolation method.

\section{The Cylindrical Cavity}
The example of a spherical cavity shows that the SCE in some instances is surprisingly accurate. However, there
evidently are systems without periodic classical orbits, such as the two perpendicular planes investigated
in\cite{GK05}, or the Casimir pendulum of\cite{SJ05}. None of these systems is integrable, and although there
are no stationary periodic classical rays,  periodic rays of \emph{extremal} (shortest) length do exist.
Semiclassically, such extremal periodic rays are associated with \emph{diffraction}\cite{Diffraction,MS04}. The
inclusion of diffractive contributions in the semiclassical estimate of Casimir energies has so far only been
attempted for a system of spheres\cite{VWR94}. It will become evident below that diffractive contributions also
play a central role in the Casimir energy of a cylindrical cavity.

The Casimir energy of a dilute cylindrical gas of atoms was found to vanish in\cite{Romeo98}. A number of
calculations have confirmed that there is no contribution up to second order in the reflection coefficients for
dielectrics\cite{Barton01,CM05,RM05} and for media where the speed of light on either side of an infinitesimally
thin cylindrical boundary is the same\cite{LNB99,BN94,MNN99,KR00,BD04}. Balian and Duplantier even conjectured
that the Casimir energy of an ideal metallic cylindrical cavity may vanish\cite{BD04} to all orders of the
multiple reflection expansion. The non-vanishing Casimir energy of an ideal metallic cylindrical
cavity\cite{DRM81} was reanalyzed in the framework of zeta-function regularization. It was confirmed that the
Casimir energy of an ideal metallic cylinder only vanishes to leading order and that higher orders in the
reflection coefficients all give a non-vanishing contribution\cite{GR98}. However, some mathematical prowess is
required to analytically prove the lowest order cancellation in the field-theoretic approach\cite{CM05,RM05}.
That a number of separate contributions should conspire to a null result without apparent physical reason has
been considered by many as somewhat "mysterious"\cite{LNB99,Milton04}. The suspicion that this cancellation
could be a purely geometrical effect, is nourished by the fact that the finite part of the pair-wise Van
DerWaals interaction energy of a dilute gas of atoms vanishes for a cylinder\cite{Barton01,Romeo98} but not for
other geometries. In\cite{MSSS03} it was found that the semiclassical contribution due to periodic rays also
vanishes for a cylinder. However, a careful perturbative analysis reveals that the interaction energy of any
real dilute cylindrical gas of atoms includes a logarithmic divergence in addition to divergent contributions
proportional to the volume and surface area of the cylinder\cite{Barton01}. The subtraction of this logarithmic
divergence generally is ambiguous and the Casimir energy of a cylindrical cavity depends sensitively on
properties of its boundary\cite{BP01} in the ultraviolet. A particular boundary (say an infinitesimally thin
cylindrical shell separating media with the same speed of light) thus may have a finite (negative) Casimir
energy, whereas a very small modification of this boundary (say finite thickness) leads to a logarithmic
divergence.

The calculation below supports this possibility. It is already known from\cite{MSSS03} that the semiclassical
contribution to the Casimir energy vanishes for a cylinder. The SCE of a cylindrical cavity in fact vanishes to
\emph{all orders} in the reflection coefficients for the same reason that the SCE of a spherical cavity is
positive -- due to relatively obvious optical phases. The semiclassical point of view thus gives a
straightforward and physically acceptable explanation for otherwise mysterious cancellations. It also indicates
that \rm{any} additional phase change at the boundary will destroy this delicate mechanism. The finite
electromagnetic Casimir energy\cite{DRM81} of a cylinder with idealized metallic boundary conditions on the
other hand is more difficult to explain semiclassically. However, contrary to a spherical cavity and in
agreement with the perturbative result of\cite{Barton01}, the semiclassical expression for the Casimir energy of
a cylindrical cavity also is logarithmic divergent. This is due to "diffractive" end-point contributions that
are ignored in stationary phase approximation. There is reason to believe\cite{Saharian00,ST06} that the
subtraction of this logarithmic divergence by the contribution from "exterior" modes gives a finite Casimir
energy for an ideal metallic and infinitesimally thin cylinder\cite{DRM81}.

To exhibit these effects we revisit the calculation of the electromagnetic SCE of a long cylindrical cavity, or
rather of a very thin torus with one perimeter $L$ that is much larger than the other, $L\gg 2\pi R$. This is an
integrable system. In the limit $R/L\rightarrow 0$, the only classical trajectories of relevance are again those
of Fig.~1 and the SCE of a long cylindrical cavity can be obtained along similar lines as that of a spherical
one -- with some important modifications. Due to the toroidal symmetry of the (long) cylinder, the third action
$I_3= L p_L/(2\pi)$ in this case is proportional to the conserved momentum $p_L$ along the axis of the (thin)
cylinder and in\equ{H} the energy $E$ must be replaced by $\sqrt{E^2-(2\pi c I_3/L)^2}$. The second action
furthermore is the angular momentum rather than just its magnitude. It again is convenient to consider
dimensionless quantities for the fraction $-1\leq x\leq 1$ of the total momentum along the axis of the cylinder,
for the ratio $-1\leq z\leq 1$ of the angular momentum to the maximal possible angular momentum of a photon
within the cavity and for its energy $0\leq\lambda<\infty$ in units of $\hbar c/(2R)$,
\bel{var1}
\lambda=2 ER/(\hbar c),\ z= \frac{c I_2/(ER)}{\sqrt{1-x^2}},\ x=\frac{2\pi c I_3}{E L}\ .
\ee
Proceeding as in the spherical case, the semiclassical expression in\equ{dualtrafo} for the SCE of a massless
scalar field satisfying Neumann or Dirichlet boundary conditions on a cylindrical surface becomes,
\bal{cylinder}
{\cal E}_{\rm cyl}\hspace{-.7em}&=&\hspace{-.7em}\frac{\hbar c L}{16\pi^2 R^2}{\sum_{m,n\geq 0}}^\prime
\Re\Big[e^{-i\frac{\pi}{2}\beta(n,m)}\hspace{-.5em}\int_0^\infty \hspace{-1em}d\lambda\hspace{-.5em}\int_{-1}^1
\hspace{-1em} dz\hspace{-.5em}
\int_{-1}^1\hspace{-1em} dx\; \times\nonumber\\
&&\hspace{-2em}\times  \lambda^3{\scriptstyle \sqrt{1-z^2}} e^{i\lambda\sqrt{1-x^2}[n
(\sqrt{1-z^2}-z\arccos(z))+m\pi z]}\Big]\
.\nonumber\\
\ea
The contribution from periodic orbits that wind around the perimeter of the torus is negligible in the
$R/L\rightarrow 0$ limit and has been omitted in\equ{cylinder}. The phase of the integrand in\equ{cylinder} is
stationary at $\bar x=0$ (corresponding to $p_L=0$) and $\zb(n,m)$ given in\equ{sp}. Since the domain of
integration for the $z$-variable differs from the spherical case, sectors with $1<m<n-1$ have non-trivial
stationary points. The classical action of an $(n,m)$-sector is the same as for the spherical cavity and is
given by\equ{Scl}. The fluctuations about such a classical ray on the other hand are quite different for
cylindrical and spherical cavities. To quadratic order in the fluctuations about the stationary point $\bar x=0,
\zb(n,m)$, the action for the cylinder is
\bel{Scyl}
S_{nm}\sim n\Big[(1-\frac{x^2}{2})\sin{\frac{m\pi}{n}}+\frac{(z-\zb(n,m))^2}{2\sin{\frac{m\pi}{n}}}\Big].\
\ee
The unconstrained Gaussian integrals over $z-\zb(n,m)$ and $x$ result in a factor of $2\pi/(n\lambda)$ in
stationary phase approximation. Note that the phases of $\pm \pi/4$ associated with the two Gaussian integrals
cancel in this case. Performing also the integral over $\lambda$ in \equ{cylinder} finally gives,
\bel{cyl1}
{\cal E}_{\rm cyl}=\frac{\hbar c L}{4\pi R^2}\sum_{n=2}^\infty \sum_{m=1}^{n-1} \Re
\frac{-i\,e^{-i\frac{\pi}{2}\beta(n,m)}}{n^4\sin^2{\frac{m\pi}{n}}}\ .
\ee
The crucial difference to a spherical cavity is the phase factor of $-i$. It arises because the fluctuations of
a cylindrical system have one fewer zero-mode than those of a spherical one\footnote{The Hamiltonian of a
spherical cavity does not depend on $I_3\propto L_z$, whereas it does depend on $I_3\propto p_L$ for the
cylindrical cavity. The $2\times 2$ Hessian matrix $H_{ij}=\partial^2 H/\partial I_i\partial I_j$ with $3>i,j>1$
has one zero mode for a spherical cavity, but none for the cylindrical geometry. This difference in zero modes
implies\cite{Schaden06} an additional phase loss of $\pi/2$ for the periodic rays of a cylindrical cavity.}. The
additional phase loss of $\pi/2$ ultimately is responsible for the vanishing of the SCE of a cylindrical cavity.
To verify this we only need to compute the Keller-Maslov index $\beta(n,m)$ for Neumann and Dirichlet boundary
conditions. The caustics of the cylindrical cavity are of first order rather than second: the cross-section of a
bundle of rays becomes one-dimensional at the caustic -- it is focussed to a line rather than a point. Taking
into account the phase retardation by $\pi/2$ every time a ray passes a first order caustic, the analogous
result to \equ{KM} for a cylindrical cavity is,
\bel{KMc}
\beta(n,m)=\left\{\begin{array}{rl} 3n, &\mathrm{for~Dirichlet~b.c.}\\
n, &\mathrm{for~Neumann~b.c.}\end{array}\right.
\ee
Contributions from paths with Neumann and Dirichlet boundary conditions and an odd number of reflections cancel
each other and, as for the spherical cavity, only sectors to even $n=2 k=2,4,\dots $ contribute to the
electromagnetic SCE [for smooth metallic cavities this is quite generally so\cite{BD04}]. Summing contributions
to the electromagnetic Casimir energy from the two scalars in \equ{cyl1} then gives the null result
\bel{cyl2}
{\cal E}^{\rm EM}_{\rm cyl}=\frac{\hbar c L}{32\pi R^2}\sum_{k=1}^\infty\sum_{m=1}^{2 k-1} \Re \frac{-i (-1)^k
}{k^4\sin^2{\frac{m\pi}{2 k}}}=0\ .
\ee

In\equ{cyl2} \emph{every} periodic orbit gives a vanishing contribution to the SCE of a cylindrical cavity. The
cancellation evidently depends on a delicate relation between the optical phases.  It is interesting that a
small additional phase \emph{loss} at each reflection off the surface results in a \emph{negative} SCE for a
cylindrical cavity, but that the Casimir energy vanishes as long as the above phase relations hold -- even if
the magnitude of the reflection coefficients is less than unity. The SCE in this sense is in line with previous
results\cite{Barton01,CM05,RM05,Romeo98} for the Casimir energy of a dilute dielectric cylinder, and in fact
supports the conjecture of Balian and Duplantier in\cite{BD04}.  The non-vanishing Casimir energy of a
cylindrical cavity with an infinitesimally thin ideal metallic boundary on the other hand is not so easily
explained by this semiclassical point of view.

Some insight is gained by noting that the contribution of \emph{any} sector to the SCE of a cylindrical cavity
in\equ{cylinder} -- even sectors with non-trivial periodic classical paths -- diverges. This is in marked
contrast to the spherical case, where the contribution from sectors with non-trivial periodic classical paths
(characterized by $n\ge 2 m>1$) is \emph{finite}. The divergence is readily made explicit by scaling
$\lambda\sqrt{1-x^2}\rightarrow\lambda$ in the integral of\equ{cylinder}. Without ultraviolet cutoff, the
resulting $x$-integral in this case formally gives the factor,
\bel{divint}
\int_{-1}^{1} \frac{dx}{(1-x^2)^2}\sim\infty\ ,
\ee
which diverges due to the behavior of the integrand near the endpoints of the integral at $x=\pm1$. It may be
regulated by introducing an \emph{ultraviolet} cutoff $\Omega$ of some sort for the energy integral [that is in
the integral over $\lambda$]. As may be seen from\equ{divint}, the regulated integral will always include terms
that are \emph{logarithmically} divergent as $\Omega\rightarrow\infty$. The subtraction of a logarithmic
divergence depends on details of the cutoff and thus is sensitive to ultraviolet properties of the
boundary\cite{CD79}. The evaluation of (in principle divergent) integrals in stationary phase could be
considered\emph{one} way of subtracting this divergence. Because the divergence is logarithmic, the subtraction
is by no means unique in this case. The presence of a logarithmic divergence for cylindrical cavities was first
emphasized by Barton\cite{Barton01} in his perturbative treatment of a dilute gas of atoms. It also is evident
in the contribution from interior modes to the Casimir energy of an ideally metallic cylinder\cite{Saharian00}.

The foregoing is compatible with previous results\cite{DRM81,MNN99,LNB99} that the Casimir energy of a
cylindrical cavity is finite if the speed of light inside and outside its \emph{infinitesimally thin} boundary
are the same. It for instance is negative for idealized metallic boundary conditions\cite{DRM81}. The Casimir
energy in this case apparently does not suffer from any logarithmic divergences (or equivalently, from any pole
ambiguities in zeta function regularization). The Casimir energy is finite for the infinitesimally thin
boundary, because the logarithmic divergent contribution from interior modes is precisely cancelled by the
similarly logarithmic divergent contribution from exterior modes. Since the boundary is infinitesimally thin and
the speed of light on both sides of the boundary is the the same, a precise cancellation is possible. The
divergence reappears for a dielectric cavity in vacuum with a lower speed of light in the dielectric\cite{BP01}.
This occurs for a spherical as well as for a cylindrical cavity, but with an important difference: the
divergence in the spherical case is not logarithmic and may be unambiguously subtracted\cite{Barton04}. The
subtraction of the logarithmic divergence in the Casimir energy of the cylindrical cavity on the other hand
requires the introduction of some energy scale that describes ultraviolet properties of the boundary. An
analogous problem would be encountered for an ideal metallic boundary of finite thickness\cite{Saharian00} and
in fact for almost any small deviation from an idealized and infinitesimally thin cylindrical boundary between
two media with identical speed of light. Paradoxically, \emph{defining} the Casimir energy of an cylindrical
cavity in a manner that does not depend on detailed ultraviolet properties of its boundary appears all but
impossible.

It perhaps is worth mentioning in this regard that the Casimir energy of a massless scalar excitation \emph{on}
the two-dimensional spherical or toroidal boundaries is well-defined. For a spherical shell and a very thin
torus, this Casimir energy has the same dependence on the dimensions as the Casimir energies of the
corresponding cavities. For a two-sphere ($S_2$) and a very thin torus $T_2$ with $L\gg 2\pi R$ these Casimir
energies are
\bal{Cassurf}
{\cal E}_{S_2} &=& 0\\
{\cal E}_{T_2} &=& -\frac{\hbar c L}{4\pi^3 R^2} \zeta(3)\sim -0.0097\dots \frac{\hbar c L}{R^2}\ .\nonumber
\ea
Note that these Casimir energies of a massless scalar on two-dimensional spherical and toroidal surfaces are
exactly reproduced semiclassically\cite{Schaden06,Dowker71}. The presence of scalar surface modes therefore does
not change the Casimir energy of a spherical cavity but could very well contribute to that of a cylindrical one.
The Casimir energy of a massless degree of freedom on a torus not only is of the same form, but also of the same
sign and order of magnitude as the Casimir energy of an ideal metallic cylindrical cavity\cite{DRM81,GR98}. Such
a contribution from massless surface modes thus might be important for a cylindrical cavity and would be
difficult to separate from the contribution due to cavity modes.

\section{Discussion}
The semiclassical approximation to the Casimir energy of a cavity to leading order includes only contributions
from quadratic fluctuations about stationary periodic classical rays. Since all periodic rays lie in the
interior, the SCE of a concave cavity to leading order depends on the exterior only indirectly through
reflection coefficients. Periodic classical rays furthermore are of finite length. Their contribution to the
Casimir energy thus is ultraviolet finite. However, this approximation is sensible only if UV-divergent
contributions to the vacuum energy can be subtracted unambiguously from the spectral density.  Logarithmically
divergent contributions to the vacuum energy require a subtraction scale\cite{CD79}. The latter is a clear
indication that the subtraction cannot be universal since it depends sensitively on the UV-properties of the
boundary. Small changes in the boundary conditions in this case do not necessarily correspond to small changes
in the Casimir energy. The \emph{local} properties of a boundary apparently include its thickness: whereas the
Casimir energy of a cylindrical cavity with an ideal and infinitesimally thin metallic boundary is
finite\cite{DRM81} to any order in the (real) reflection coefficients\cite{GR98}, a logarithmic dependence on
the cutoff appears in more realistic situations\cite{Barton01,Saharian00}. An ambiguous subtraction also is
required in the semiclassical approximation. The absence of any logarithmic divergence for the infinitesimally
thin boundary apparently is due to a cancellation by exterior modes. Such a cancellation of logarithmic
singularities can occur when exterior and interior modes depend on precisely the same scale, the radius $R$ of
the cylindrical cavity in this case. Although the two logarithmic divergences (each proportional to $\hbar c
L/R^2$ for dimensional reasons) cancel in the idealized situation, they would not do so for a boundary of finite
thickness.

We considered only the semiclassical Casimir energy (SCE) of a spherical and of a toroidal cavity with ideal
metallic boundary conditions, that is, with real reflection coefficients of unit magnitude. These are integrable
systems and the SCE was obtained from the "dual" description of the spectral density in terms of periodic paths
on invariant tori\cite{Gutzwiller90,BT76}. The winding numbers of a periodic orbit are dual to the quantum
numbers of a mode.  In stationary phase approximation the SCE of a spherical cavity is \emph{positive} and
reproduces the field theoretic value for an infinitesimally thin metallic boundary to within 1\%. The
calculation is rather short and straightforward and leads to the convergent sum of\equ{sphere1}. Each term in
this sum may be interpreted as the contribution from a class of periodic rays. A few of the shorter primitive
periodic rays are depicted in Fig.~1. The contribution from any sector is finite in this case.

The contribution from periodic orbits to the SCE of a cylindrical cavity with an ideal metallic boundary on the
other hand vanishes to all orders in the number of reflections\cite{MSSS03}. This occurs due to an overall phase
change by an odd multiple of $\pi/2$ for any classical periodic ray. Restricting to just two reflections, this
null result agrees with field theoretic calculations for infinitesimally thin metallic
boundaries\cite{LNB99,MNN99,NP00,BD04,NP99}. The vanishing SCE appears to support the conjecture of Balian and
Duplantier that the Casimir energy of a metallic cylindrical cavity may vanish. However, contrary to the
spherical case, the contributions of \emph{any} classical sector to the SCE of a cylindrical cavity diverges.
Without subtraction of the UV-divergent part, the (finite) semi-classical contribution to the vacuum energy we
obtained is not very meaningful. Unfortunately the divergence of the integral in\equ{divint} includes a
logarithmic dependence on the cutoff. The subtraction of UV-divergent contributions to the Casimir energy of a
cylinder thus is sensitive to a scale and cannot be achieved in a universal fashion. The logarithmic dependence
on the cutoff was first observed by Barton\cite{Barton01} in his perturbative calculation of the vacuum energy
for a (dilute) gas of cylindrical shape to lowest order in the fine structure constant. Semiclassically this
would also correspond to considering the contribution from rays with only two reflections ($n=2$). That the
UV-subtractions are fragile and depend crucially on the UV-properties of the boundary is also observed when the
speed of light within and outside an infinitesimally thin cylindrical boundary differ\cite{BP01}. In the
electromagnetic case, the logarithmic divergences of exterior and interior contributions to the vacuum energy of
a metallic cylinder cancel for an infinitesimally thin metallic boundary\cite{Saharian00}. However, they in
general cannot be unambiguously subtracted\cite{ST06}.

These examples of a spherical- and cylindrical cavity show that the SCE is quite reasonable and is rather simple
to calculate when the Casimir energy is robust, that is, when the necessary subtractions depend on global
properties of the system only and do not require the introduction of an additional scale. This apparently is not
possible if the divergence is logarithmic\cite{CD79,Schaden06}. The classical periodic paths that contribute to
the SCE of a concave cavity in stationary phase approximation lie entirely within the cavity. Their contribution
depends on the exterior of the cavity through reflection coefficients only.  It has been argued for some time
that a Casimir energy obtained without explicit inclusion of exterior modes (as for a
parallelepiped\cite{Lukosz71,AW83,Edery06}) is all but meaningless\cite{Milton04}. The criterion favored
here\cite{Barton01,Schaden06} considers any definition of a Casimir energy reasonable (and in principle
physically realizable) in which the UV-divergences of the vacuum energy have been subtracted without reference
to local properties of the boundary. The subtraction may (and in general will) include divergent contributions
from exterior modes. The Casimir energy of a parallelepiped can be considered a case in point: as
Power\cite{Power64} did for just two slabs, one can always assemble parallelepipeds to a cube with fixed
dimensions -- the Casimir energy of an individual parallelepiped\cite{Lukosz71,AW83,Edery06} in this case
reflects changes in the vacuum energy of the whole cube as the three dividing planes are moved adiabatically. By
moving interior surfaces of the cube (that in principle could have finite thickness), one measures only the
finite part of its vacuum energy that depends on the dimensions of the individual parallelepipeds. By contrast,
it is difficult to imagine that global changes in a vacuum energy are measurable (or even physically relevant)
if their finiteness depends critically on local characteristics of the system\cite{Schaden06}. Perhaps somewhat
surprisingly, the electromagnetic Casimir energy due to the interior of a very long cylindrical cavity does not
appear to be robust in this sense, whereas the electromagnetic Casimir energy due to the interior of a spherical
cavity is.

Apart from relating Casimir energies to optical properties, one of the advantages of a semiclassical description
would be the possibility to model more realistic (but robust) physical systems. The previous considerations are
readily extended to dielectrics by using appropriate complex and in general frequency-dependent reflection
coefficients. In the case of dielectric slabs Milton has shown\cite{Milton04} that Lifshitz's
theory\cite{Lifshitz56,Milloni93} may be reproduced in this manner. Finite temperature is
incorporated\cite{temperature} by allowing periodic rays to also wrap around a fictitious periodic extra
dimension of circumference $\hbar c/(kT)$. Finite temperature corrections thus are small if some classical
periodic paths are much shorter than this circumference. At room temperature the length of a periodic ray
increases by about $7.6$ microns every time it winds about the temperature direction. Temperature corrections
therefore are tiny for most nanometer scale experiments\footnote{The claim that this correction has been
measured to sufficient accuracy\cite{Experiments99} to distinguish between different approaches has recently
been disputed\cite{HBAM06}.} but could be of greater interest in some astrophysical considerations ($3^o K\equiv
1$mm). Corrections due to surface roughness generally will be more important in technological applications. Many
classical models for diffuse reflection from rough surfaces exist and Lambert's Law is easily incorporated in
the semiclassical approach by appropriate reflection coefficients. The dependence on the wavelength perhaps can
be modelled by a term of the action that accounts for stochastic fluctuations in the length of a classical
periodic orbit upon reflection from rough surfaces. Apart from an average change in length, this leads to a
damping term of the form $-(\Delta L\, E/\hbar c)^2/2$ in the classical action, where $(\Delta L)^2$ is the
variance in the length of the periodic orbit. Assuming that this variance is itself proportional to the length
of the orbit, surface roughness can semiclassically perhaps be modelled by the modified dispersion
\bel{roughdisp}
c p(E) = E +i \eps E^2/(\hbar c)\ ,
\ee
where $\eps$ is a typical length scale for the (stochastic) roughness of the surface. The predominant effect of
the modified dispersion of\equ{roughdisp} is that contributions to the Casimir energy from wave lengths
$\lambda\ll\eps$ are very much suppressed. A similar conclusion may be drawn from a recent and considerably more
sophisticated analysis\cite{NLR05}.

{\bf Acknowledgements:} I am indebted to Larry Spruch for advice and numerous discussions and would like to
thank K.~A.~Milton for a critical reading of the manuscript. This work was partially supported by the National
Science Foundation under Grant No.~PHY0555580.

\end{document}